\documentclass[letterpaper,aps,twocolumn,superscriptaddres,nofootinbib]{revtex4-2}
\usepackage{graphicx, color, amsmath,amssymb,latexsym,cleveref, multirow, adjustbox}

\def\lsim{\mathrel{\raise.3ex\hbox{$<$\kern-.75em\lower1ex\hbox{$\sim$}}}}
\def\gsim{\mathrel{\raise.3ex\hbox{$>$\kern-.75em\lower1ex\hbox{$\sim$}}}}

\begin{document}

\title{Effects of large-scale magnetic fields 
  on the observed composition \\ 
  of ultra high-energy cosmic rays }

\author{Ellis R. Owen}
\email{erowen@astro-osaka.jp}
\thanks{JSPS International Research Fellow}
\affiliation{Theoretical Astrophysics, Department of Earth and Space Science, Graduate School of Science, Osaka University, Toyonaka, Osaka 560-0043, Japan}
\affiliation{Institute of Astronomy, National Tsing Hua University, Hsinchu, Taiwan (ROC)}
\affiliation{Center for Informatics and Computation in Astronomy, National Tsing Hua University, Hsinchu, Taiwan (ROC)} 
\author{Qin Han}%
\affiliation{Mullard Space Science Laboratory, University College London, Holmbury St Mary, Surrey RH5 6NT, UK
}%
\author{Kinwah Wu}
\affiliation{Mullard Space Science Laboratory, University College London, Holmbury St Mary, Surrey RH5 6NT, UK
}

\date{\today}

\begin{abstract}
Ultra high-energy (UHE) cosmic rays (CRs) 
 from distant sources 
 interact with intergalactic radiation fields, 
 leading to their spallation and attenuation. 
They are also deflected in 
  intergalactic magnetic fields (IGMFs), 
particularly those associated with Mpc-scale structures.  
 These deflections 
  extend the propagation times of CR particles, forming a magnetic horizon for each CR species. 
  The cumulative cooling and interactions of 
   a CR ensemble also 
  modifies their spectral shape and composition observed on Earth.
We construct a transport formulation 
 to calculate the observed UHE CR spectral composition 
 for 4 classes of source population. 
The effects on CR propagation brought about by IGMFs 
  are modeled as scattering processes during transport, 
  by centers associated with cosmic filaments. 
Our calculations demonstrate that IGMFs 
  can have a marked effect 
  on observed UHE CRs, 
 and that source population models  
 are degenerate with IGMF properties. 
Interpretation of observations, 
including the endorsement or rejection of 
any particular source classes, 
thus needs careful consideration of 
the structural properties  
  and evolution of IGMFs. 
Future observations providing tighter constraints  
  on IGMF properties 
  will significantly improve confidence  
  in assessing UHE CR sources
  and their intrinsic CR production properties. 
\end{abstract}


\maketitle

\section{Introduction}
\label{sec:introduction}

Ultra high-energy (UHE)\footnote{We adopt the terminology that cosmic rays 
  with energies above $10^{17}\;\!{\rm eV}$ 
  are referred as UHE cosmic rays \cite[e.g.][]{Kachelriess2019PrPNP}.} 
  cosmic-rays (CRs) 
  are believed to originate in violent astrophysical environments, 
  e.g. blazar jets, strongly magnetized 
  neutron stars and starburst galaxies 
  \citep[see e.g.][]{Anchordoqui2019PhR}. 
Their detection on Earth is rare  
  \cite{Kachelriess2019PrPNP}, 
  with arrival rates of about $1\;\!{\rm km}^{-2}\;{\rm yr}^{-1}$
  being typical for particles with energies $E = 10^{19}~{\rm eV}$ \citep[see e.g.][]{Watson2014RPPh,Alves2019FrASS,Aab2020PhRvD}. 
UHE CRs interact with baryons and photons\footnote{These photons 
  are mainly contributed by the cosmological microwave background 
  (CMB). Extra-galactic background light (EBL) can also have some effect~\cite[see][]{Aloisio2013APh_b}.}  
  as they propagate through intergalactic space.   
 CR nuclei are converted to lighter particles 
via processes 
  such as photo-spallation and photo-pion production. 
These attenuate CR fluxes 
  and limit the survival distance of individual CRs. 
In the present Universe, 
  CR protons with energies 
  of $\sim 10^{19}\;\!{\rm eV}$ 
    {would undergo a photo-attenuation interaction over} 
  a few tens of Mpc  
  \citep[see e.g.][]{Dermer2009herb}.  
It has therefore been argued 
  that extragalactic UHE CRs  
  of energies above $\sim 10^{19}\;\!{\rm eV}$
  detected on Earth 
  may originate from discernible sources 
  {within a photo-pion horizon distance of a few tens of Mpc}  
  \citep[see][]{Greisen1966PhRvL,Zatsepin1966JETPL}. 
This is a reference distance, 
  above which the Universe 
  becomes `optically thick' to UHE CR photo-pion attenuation. 
  
{UHE CRs could originate from more distant source populations located beyond {their photo-pion horizon}~\cite[e.g.][]{Berezinskii1988AA, Owen2021ApJ}}. Their residual `background' contribution 
 can be 
 used to study possible 
 source population distributions over redshift because the cumulative effect of 
 photo-spallation as the CR ensemble propagates 
 modifies its arrival composition 
 on Earth. 
 This is also dependent on 
 the effective travel-distances  
  of UHE CR protons and nuclei, which 
  are altered 
  when magnetic fields are present 
  \cite[e.g.][]{Kotera2008PhRvD, Taylor2009PhRvD, Capel2019MNRAS, Vliet2022MNRAS}. 
 Magnetic fields permeate intergalactic space.   
  They are highly non-uniform,  
  and could have fractal structures  
  if associated with turbulence 
  \citep[see e.g.][]{Durrer2013AARv}.  
UHE CRs in intergalactic space are therefore not free streaming, 
  nor do they gyrate 
  around an ordered large-scale magnetic field.  
They are deflected 
  in a stochastic manner 
  \citep[see e.g.][]{Giacalone1999ApJ,Yan2008ApJ,Kotera2008PhRvD,Xu2013ApJ} and may also undergo diffusion 
  \cite{Kotera2008PhRvD3,Mollerach2013JCAP,Gonzalez2021PhRvD}.

 Previous studies \citep[see][]{Alves2019FrASS} 
   investigated
   UHE CR composition and spectra 
  the presence of 
  intergalactic magnetic fields (IGMFs).\footnote{We use IGMFs to refer to all large-scale magnetic fields hereafter.} 
  Ref. \cite{AlvesBatista2017PhRvD} invoked 4-dimensional simulations 
  to investigate the composition and anisotropy of UHE CRs with IGMFs 
  at different positions in a simulation box when considering CR injection of $^{56}$Fe or $^{1}$H. 
 Ref. \cite{Wittkowski2017PoS} extended this to use a more physically-motivated UHE CR source composition.  
Other studies 
  considered the effect 
  of a magnetic field structure 
  similar to the Local Super-cluster 
  on CRs 
  observed on Earth, and assessed the energy dependence of 
  CR flux suppression 
  caused by photo-pion attenuation and magnetic horizons~\cite{Mollerach2013JCAP}. 
  Although large scale inhomogeneities 
  from structures such as cosmic filaments and voids 
  were not explicitly considered, 
  secondary particle production
 was recently  
 added as a refinement 
 \citep[see][]{Gonzalez2021PhRvD}.  

In this paper, we assess the effect of IGMFs on the spectrum and composition of UHE CRs. We consider different CR source populations and magnetic field 
  prescriptions. 
To our knowledge, this study is the first to model 
  CR propagation in IGMFs with inhomogeneities over cosmological scales, 
  while properly accounting for photo-pion (absorption) and photo-spallation processes. 
It is also the first 
  to assess the effects of an inhomogeneous IGMF 
  on CR propagation 
  and whether or not it can unambiguously be   
  discerned in the observed CR spectrum 
  and composition on Earth. 
The assumptions and methodology  
 in our previous work \citep{Owen2021ApJ} \footnote{ Throughout this work, 
  we use dimensionless energies as in \cite{Owen2021ApJ}, defined in terms of electron rest mass $\epsilon=E/m_{\rm e} c^2$. }
  are adopted here,  
 {with, in addition, the 
  treatment}  of the effects brought about by IGMFs. 

We arrange this paper as follows.  
{Section}~\ref{sec:source_model}  
   introduces CR source population models, compositions and their spectra. 
{Section}~\ref{sec:UHECR_interactions} 
     presents our treatment 
    of UHE CR interactions. 
    We introduce our demonstrative magnetic field prescriptions in {Section}~\ref{sec:magnetic_field_models}. 
 {Section}~\ref{sec:results}   
   shows our results and 
   discusses their implications.  
 A summary of our findings is provided in {Section}~\ref{sec:conclusions}. 

\begin{table*}
\begin{tabular}{lcr}
 \hline
 Parameter & Definition & Value \\ 
  \hline
 $n_c$ & Comoving number density of scattering centers & $10^{-2}$ Mpc$^{-3}$ \\
 $\sigma_c$ & Effective scattering center cross sectional size & 3 Mpc$^{2}$ \\
  $r_c$ & Characteristic diameter of scattering center & 2 Mpc \\
  $\lambda_c$ & Magnetic field coherence length within the scattering structure & 0.3 Mpc \\
  \hline
  \end{tabular}
 \label{tab:mag_list}
\caption{Fiducial parameter choices for our illustrative  magnetic field prescription. This considers cosmic filaments as the dominant 
scattering agent. Filament properties are based on the prescription introduced by~\cite{Kotera2008PhRvD}. The magnetic field strength evolves over redshift (see Fig.~\ref{fig:z_evo_magfield}).}
\end{table*}
   
\section{Cosmic ray sources, propagation and interactions}
\label{sec:cr_model}

\subsection{Source population models} 
\label{sec:source_model}

We consider four UHE CR source population models, 
  specified over a redshift range 
  from $z_{\rm min} = 0$ to $z_{\rm max} = 3$. 
For each model, the source number density 
  in redshift space, 
  composition and 
  injected CR energy spectra follows the same paramatrization as in~\cite{Owen2021ApJ}. 
We summarize the source population models  
  and the corresponding paramatrizations 
  as follows. 
The first is the SFR model. It follows the redshift evolution of cosmic star formation (see also~\cite{Muzio2019arXiv}) {and takes the form:}
\begin{equation}
\label{eq:z_SFR}
    {\psi_{\rm SFR}(z) = \psi_{\rm SFR}^{0} \;\! \frac{(1+z)^{k_1}}{1+[(1+z)/k_2]^{k_3}} \ ,}
\end{equation}
  {where $k_1 = 2.7$, $k_2 = 2.9$, $k_3 = 5.6$ and $\psi_{\rm SFR}^{0} = 0.054$.}
The second is the GRB model. 
This is an adjustment of the SFR model and represents a possible redshift distribution of gamma-ray bursts. 
Its construction is based on \textit{Swift} observations 
that indicate a similar redshift distribution to cosmic star-formation, but with an enhancement at earlier epochs. 
{For the GRB model, 
we consider a redshift distribution,} 
\begin{equation}
\label{eq:z_GRB}
    {\psi_{\rm GRB}(z) = \psi_{\rm GRB}^{0} \;\! (1+z)^{k_4} \;\! \psi_{\rm SFR}(z) \ , }
\end{equation}
{where $k_4 = 1.4$ and $\psi_{\rm GRB}^{0} = 0.013$,  
     following \cite{Wang2011ApJ}. }
The third is the AGN model.  {
We adopt an AGN population evolution parametrization, 
  given by~\cite{Hasinger2005AA}:}
 \begin{equation}
\label{eq:z_AGN}
    {\psi_{\rm AGN}(z) = \psi_{\rm AGN}^{0} 
\left\{
    \begin{array}{ll}
        (1+z)^{k_5} \hspace{0.5cm}  & (z < z_1)\  \\
        {z_2}^{k_5} \hspace{0.5cm} & (z_1 \leq z < z_2)\  \\
        {z_2}^{k_5}\cdot {z_2}^{z_2-z} \hspace{0.5cm} & (z\geq z_2)\   
    \end{array} 
\right. \ , }
\end{equation}  
  {where $k_5 = 5.0$, $z_1=1.7$, $z_2=2.7$ and $\psi_{\rm AGN}^{0} = 0.0041$. }
The fourth is the PLW model. 
It parameterizes the source population 
 by a power-law distribution in redshift space{, as:}
 \begin{equation}
\label{eq:z_PLW}
    {\psi_{\rm PLW}(z) = \psi_{\rm PLW}^{0} 
    \;\! (1+z)^{k_{\rm PLW}}  \ , }
\end{equation} 
 {with $k_{\rm PLW} = -1.6$ and $\psi_{\rm PLW}^{0} = 1.1$.}
This model is not specifically based 
  on observations or a survey. 
It serves instead as a generic basis 
  for comparison with similar PLW-type models 
  that are employed in some other studies  
  \cite[e.g.][]{Taylor2015PhRvD, AlvesBatista2019JCAP}. 

{Here, the same spectral forms 
 as in~\cite{Owen2021ApJ} 
 are adopted for UHE CRs (see their Table 1).}  
The overall injection luminosity in each model is normalized 
  by gauging against Pierre Auger Observatory (PAO) data without detailed fitting. 
The energy range of the spectra is 
  between $\epsilon_{\mathrm{min}} m_{\rm e} c^2 = 3.98 \times 10^{18}\;\!{\rm eV}$ and $\epsilon_{\mathrm{max}} m_{\rm e} c^2 = 3.16\times 10^{20}\;\!{\rm eV}$. 
This is chosen 
 to cover the CR flux contributed mostly 
 by extragalactic particles 
 \cite{Giacinti2012JCAP, Aloisio2014JCAP},   
 and extends up to the most energetic UHE CRs 
 expected to be detected on Earth~\cite{Bird1995ApJ}.

In each source class 
 the full range of injected nuclei 
 are represented by 
 the abundances 
 of $^{28}$Si, $^{14}$N, $^{4}$He, and $^{1}$H.
The injected composition fractions  
  follow the fitted values of the species  
  given in \cite{AlvesBatista2019JCAP}. 
 Variation of these fixed parameters would lead to a larger number of calculations and introduce more uncertainty, but would not improve the accuracy of our results. 
In our calculations, 
  the production of all secondary nuclei species 
  of mass number $A<28$  
  are properly accounted for. 
This safeguards 
the correct determination of 
photo-spallation interactions and their secondary products 
along particle propagations 
 (see Section~\ref{sec:UHECR_interactions}).

\subsection{Propagation and interactions of ultra high-energy cosmic ray nuclei} 
\label{sec:UHECR_interactions}

UHE CR nuclei are subject to hadronic interactions 
  and energy losses as they propagate through intergalactic space. 
A CR particle may lose only a small fraction of its energy 
 in a single interaction event, or it 
 may lose energy continuously 
  (such processes include photo-pair production, Compton scattering,   
  radiative losses, 
  or adiabatic energy losses in an expanding volume 
  or space-time). 
We model these 
  as effective ``cooling'' processes.   
In some situations, a CR particle loses 
  substantial energy 
  in a single interaction (e.g. photo-pion production) or 
can be split (photo-spallation). 
We treat these as absorption processes. 
In the case of photo-spallation events, 
 we self-consistently account for the production of descendant nuclei using appropriate injection terms.

In the absence of IGMFs, 
  the propagation of UHE CRs across intergalactic space 
  is practically ballistic streaming at $c$, the speed of light.  
  The corresponding CR transport equation is 
\begin{align}
\label{eq:transport_equation_steadystate}
\frac{\partial\;\! {n_A}}{\partial z} 
 & =  \frac{{\rm d} s}{c\;\! {\rm d} z}
    \left[\frac{\partial}{\partial \epsilon_A} 
    \left(b_A n_A \right) + {Q}_A -  \Lambda_A n_A \right]
    \ ,    
\end{align} 
when adopting a quasi-steady condition~\cite{Owen2021ApJ}.
Here, the particle species is specified by mass number $A$.
 $n_A(\epsilon_A, z)$ is the comoving spectral density of UHE CRs with mass number $A$ and dimensionless energy $\epsilon_A$, 
 $b_A$ is the total energy loss rate experienced by those CRs due to cooling processes 
 and 
 $\Lambda_A = {\Lambda^{\rm sp}}_{\!A} + {\Lambda^{A \pi}}_{\! A}$ is the total absorption rate accounting for photo-spallation (${\Lambda^{\rm sp}}_{\!A}$) and photo-pion production (${\Lambda^{A \pi}}_{\! A}$). 
 $Q_A=Q_A^{\rm a}+Q_A^{\rm sp}$ is the injection rate of UHE CRs. This is the sum of photo-spallation products (secondary nuclei) $Q_A^{\rm sp}$ and fresh primary particle injection $Q_A^{\rm a}$ by the source population. 
In a Friedmann-Lema{\^i}tre-Robertson-Walker (FLRW) universe,
\begin{align}
\label{eq:dsdz}
 \frac{{\rm d}s}{{c\;\! \rm d}z} 
 &  = \frac{\mathcal{E}(z)}{H_0\;\!(1+z)} \ ,
\end{align}
  where  $H_0$ is the present value of the Hubble parameter. This takes a value of $100\;\! h\;\! {\rm km}\;\!{\rm s}^{-1}{\rm Mpc}^{-1}$,
  $h = 0.673\pm 0.006$ 
  \citep{Planck2020AA}, and 
\begin{align}
\label{eq:e_func}
    \mathcal{E}(z) = \left[\Omega_{\rm r, 0}(1+z)^4 + \Omega_{\rm m, 0}(1+z)^3 + \Omega_{\rm \Lambda, 0} \right]^{-1/2}
\end{align}
\citep[see][]{Peacock1999book},  
  with $\Omega_{\rm m,0} = 0.315\pm0.007$, 
  $\Omega_{\rm r,0} \approx 0$ 
  and $\Omega_{\rm \Lambda,0} = 0.685\pm0.007$. 
  These are the normalized density parameters for matter, 
  radiation and dark energy, respectively~\citep{Planck2020AA}.

  If accommodating IGMFs into our formulation, CR transport is modified across intergalactic space. 
 IGMFs have certain structures associated 
  with density distributions, 
  which may be in the form of galaxies, galaxy groups/clusters 
  and cosmic filaments.  
   They could also be turbulent in nature. 
A comprehensive treatment 
  of UHE CR propagation, 
  properly accounting for 
  the effects of magnetic fields 
  convolved with those of density structures 
  of objects across the mass hierarchy in the Universe is non-trivial.    
This subject has been addressed in previous studies, 
  which provide insights into 
    transitions to diffusive propagation in detailed configurations of turbulence 
    \citep[e.g.][]{Ahlers2014PhRvL, AlvesBatista2014JCAP, Supanitsky2021JCAP}, including the effects of magnetic intermittency~\cite{Shukurov2017ApJ}, and by invoking 
 increasingly detailed simulations~\cite[e.g.][]{AlvesBatista2017PhRvD, AramburoGarcia2021PhRvD}. 
 
We consider that 
  CR transport in localized regions of enhanced magnetic fields
  may be treated as a series of discrete scattering events. 
  Hereafter, we refer to scattering events that lead to the deflection of CR particles as `deflections'. 
Magnetized regions are associated 
  with particular astrophysical environments which act as scattering centers 
  (e.g. galaxy clusters or cosmic filaments; see~\cite{MedinaTanco2001APh, Kotera2008PhRvD}). 
This heuristic treatment 
  captures the main essence 
  of UHE CR transport in intergalactic space 
  in the presence of IGMFs when the accumulated deflection angle of the CRs is small.  
It is justified, 
   as the linear sizes of the scattering structures 
   (e.g. cosmic filaments, which would be a few Mpc~\citep[e.g.][]{Doroshkevich2001MNRAS, Cautun2014MNRAS}) 
   are much smaller than 
   either the spacing between structures 
   ($\sim$ 100 Mpc~\cite[e.g.][]{Libeskind2018MNRAS}), 
   or the photo-pion horizon scale of the CR particles.

Our formulation is essentially 1-dimensional (1D) 
  and 
    the effects of deflections are captured in an  
  effective difference in path length.  
The additional path length of weakly scattered particles (with a deflection angle much smaller than $\pi/2$)
compared to free streaming CR propagation 
  is expressed as  
\begin{align}
\label{eq:extra_in_step}
    \delta s' & = N_c \;\! {\delta t_c'} \;\! c \ . 
\end{align}
Here, $N_c$ is the number of scattering events a CR undergoes along a path.  
In an interval $\delta s$, this 
 is given by $N_c = \delta s / d_c$, 
 where $d_c = (n_c \;\!\sigma_c)^{-1}$ is the mean free path of a CR to an interaction with a scattering center. $n_c$ and $\sigma_c$ are the comoving number density and physical effective cross sectional size of the scattering centers, respectively. 
 The extra propagation time $\delta t_c'$, introduced when a CR is deflected into an longer, non-rectilinear path by a scattering event, plus the delay time it experiences when crossing each scattering center, can be expressed as 
\begin{align}
\delta t_{c}' & \simeq \frac{d_{\rm c} \;\! \delta {\theta_c}^{2}}{8 c} + 
\frac{\bar{r}_{c}\;\! \delta {\theta_c}^{2}}{6\;\! c}  \  
\end{align}
\cite{Alcock1978ApJ, Kotera2008PhRvD}.  
The first term above accounts for the time delay associated with the non-rectilinear trajectory arising from the scattering event. The second term accounts for the time delay with respect to straight line when a CR crosses a magnetized scattering structure.   
$\bar{r}_{c}$ is the characteristic 
 path length through a scattering center. This is related to the physical size of the scattering center $r_c$ by $\bar{r}_c=(\pi/2)^2\;\!r_c$ (for a filamentary morphology~\cite{Kotera2008PhRvD}). 
The deflection angle 
  $\delta \theta_{c}$, is given by   
\begin{align}
 \label{eq:deflection_angle_general}
\delta {\theta_c}^{2}  & 
\simeq\left(1+\frac{2\;\! {r_L}^{2}}{\bar{r}_c \lambda_{c}}\right)^{-1} \ , 
\end{align}
where $\lambda_c$ is the coherence length of the magnetic field of the scattering structure. 
$r_{L}$ is introduced as the charged particle's gyro-radius, which depends on the CR charge $Z_A$, energy $\epsilon_A$ and the magnetic field strength
of the scattering center:  
\begin{align}
r_{L}  & \approx 1.1
\;\!\left(\frac{\epsilon_A}{2\times10^{12}}\right) \left(\frac{B}{1\;\!{\rm nG}}\right)^{-1} \left(\frac{Z_A}{1}\right)^{-1} {\rm Mpc} \ .
\end{align} 
Equation~\ref{eq:deflection_angle_general} 
  is adopted 
  for the transition from $r_L \ll \sqrt{r_c \lambda_c}$ 
  (when the particle exits a scattering center with a small deflection angle after undergoing a single scattering event) 
  to $r_L \gg \sqrt{r_c \lambda_c}$ 
  (where the particle diffuses \textit{within} the scattering structure, and exits roughly at the same location it entered; see~\cite{Kotera2008PhRvD}).
  The overall path length 
   taking account 
   of the effects of 
 deflections is then given by 
   $\delta s' = \delta s + c \delta t_c'$.

For the 
  propagation of some CRs, the accumulated deflection angle 
  can become large ($>\pi/2$). 
The CR transport in this situation 
  is instead modeled 
  using a strong deflection prescription,  
 where the additional path lengths experienced by CRs are
   given by 
   $\delta s'' = \delta s' (1 + \Delta s'/\ell_{\rm scatt})$. 
   Here, $\Delta s'$ is the integrated path length, and 
$\ell_{\rm scatt} = (n_c\;\! \sigma_c \;\! {\delta \theta_c}^2)^{-1}$ 
is the scattering distance over which the CRs experience a strong deflection.
  With 
    the additional path lengths 
    properly incorporated 
    in the calculations 
    of photo-pion interactions 
    and photo-spallation, 
  we obtain a revised relation   
    for the CR travel time and propagation distance.

With equation~\ref{eq:extra_in_step},  
  we construct a scaling factor 
  for the fractional path length extension experienced 
  by the propagating CRs due to deflections: 
\begin{align} 
\label{eq:modxx}
 \Psi_A 
 =\frac{\delta s''}{\delta s} = \frac{\delta s' (1 + \Delta s'/\ell_{\rm scatt})}{\delta s} 
\end{align} 

Applying this to the transport equation (equation~\ref{eq:transport_equation_steadystate}), 
 we obtain 
\begin{align} 
\label{eq:transport_equation_steadystate_mod}
 \frac{\partial\;\! {n_A}}{\partial z} 
  =   \frac{{\rm d} s}{c\;\! {\rm d} z}
  \bigg( 
  \Psi_A
  \bigg[ \frac{\partial}{\partial \epsilon_A} 
    ( b_A n_A & )  + \phi_{\rm A}^{B}\;\! Q_A^{\rm sp} -  \Lambda_A n_A \bigg] + {Q}_A^{\rm a} \bigg)  \ .  
\end{align}  
Note that this omits an explicit treatment of diffusion, with its effects considered only in the path length scaling term $\Psi_{\rm A}$. 
This is justified 
 in the computation of the total diffuse average fluxes   
   when the   
  overall CR deflection angles accumulated over their total path length 
  are sufficiently small, 
  or when the separation between sources is shorter 
 than both the diffusion length and the CR energy loss distance~\cite{Aloisio2004ApJ, Kotera2008PhRvD}.

The magnetic scaling term is applied to all propagation, interaction and cooling terms. 
The primary source term ${Q}_A^{\rm a}$ 
    is independent of the magnetic fields. 
It is distinct from  
  another source term $Q_A^{\rm sp}$. 
  This depends on the photo-spallation rate 
  of parent CRs of higher nucleon number. 
  The propagation of the parent nuclei are modified by 
  a scaling factor appropriate for their species, $\Psi_{\rm B}$. 
  This is related to $\Psi_{\rm A}$ by the ratio of the squared deflection angles of the parent and secondary CRs, denoted $\phi_A^B$.

\begin{figure}
    \centering
    \includegraphics[width=\columnwidth]{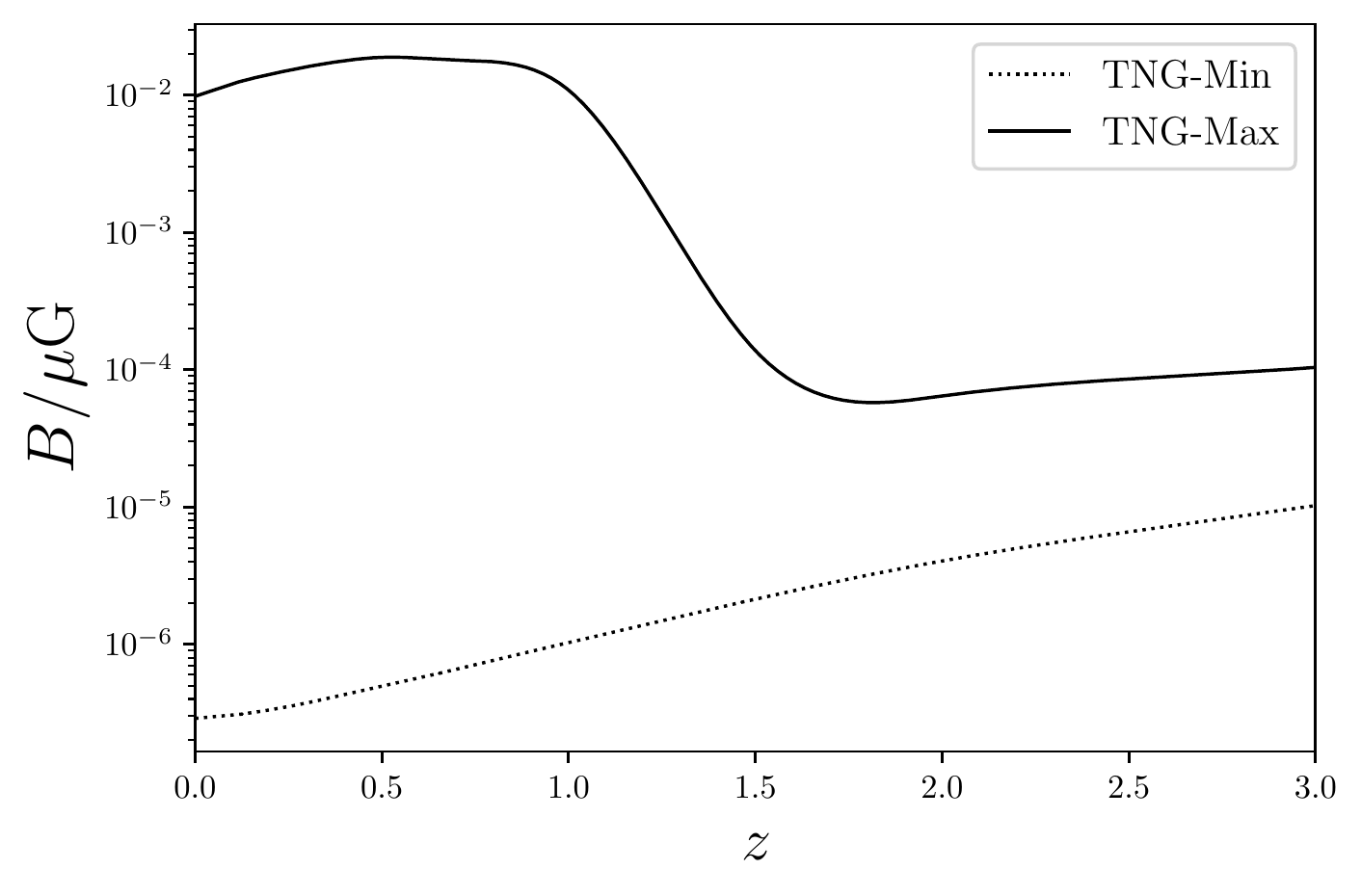}
    \caption{Evolution of magnetic field strength over redshift $z$ 
    derived from the IllustrisTNG 100-3 simulations 
    \cite{Marinacci2018MNRAS, Nelson2019ComAC}. An interpolation routine is used 
  for cosmological epochs between 
  simulation redshift slices. 
    The TNG-Max prescription is the median magnetic field strength in structures 
    with a baryon over-density with $\log_{10}(\rho/\langle \rho_b \rangle) = 2$;  TNG-Min 
    is for $\log_{10}(\rho/\langle \rho_b \rangle) = 0.7$. 
    This 
    covers the range of densities  
    appropriate for cosmic filaments~\citep[][]{Marinacci2018MNRAS}.}
    \label{fig:z_evo_magfield}
\end{figure}

\subsection{Intergalactic magnetic field prescriptions} 
\label{sec:magnetic_field_models}

The fiducial magnetic field configuration  
  is adapted from \cite{Kotera2008PhRvD}.    
The effective size, number density 
  and magnetic-field coherence length 
  of the scattering centers 
  are chosen such that 
  they are appropriate for cosmic filaments 
  (Table I).    
The evolution of the filaments 
  is not considered explicitly, 
  as their size 
  does not change substantially over 
  in the redshifts considered in this study 
  \cite{Cautun2014MNRAS}. 
The magnetic fields in the filaments, however, evolve, 
  as shown in numerical simulations 
  \cite[see e.g.][]{2014MNRASVazza}. 
We use the results from IllustrisTNG 100-3 simulations  
  \cite{Marinacci2018MNRAS, Nelson2019ComAC, Naiman2018MNRAS, Nelson2018MNRAS, Springel2018MNRAS, Pillepich2018MNRAS}) 
  to construct two redshift-dependent IGMF-strength prescriptions  
  (presented in Fig.~\ref{fig:z_evo_magfield}).

The effects of IGMFs on the propagation of UHE CRs 
  with energies between $10^{17}$~eV and $10^{21}$~eV 
  by deflections in cosmic filaments 
  can be comparable to  
  magnetized structures on galactic scales, 
  in particular, fossil radio galaxies and galactic winds 
  \cite{Kotera2008PhRvD}. 
In our calculations,   
  these substructures are incorporated implicitly 
  in the effective scattering of the filaments 
  through a maximum and minimum limit, 
  which brackets the extent of their effects.

\section{Results and discussion} 
\label{sec:results}

\subsection{UHE CR spectrum and composition} 
\label{sec:results_flux_comp}

\begin{figure*}
    \centering
    \includegraphics[width=0.95\textwidth]{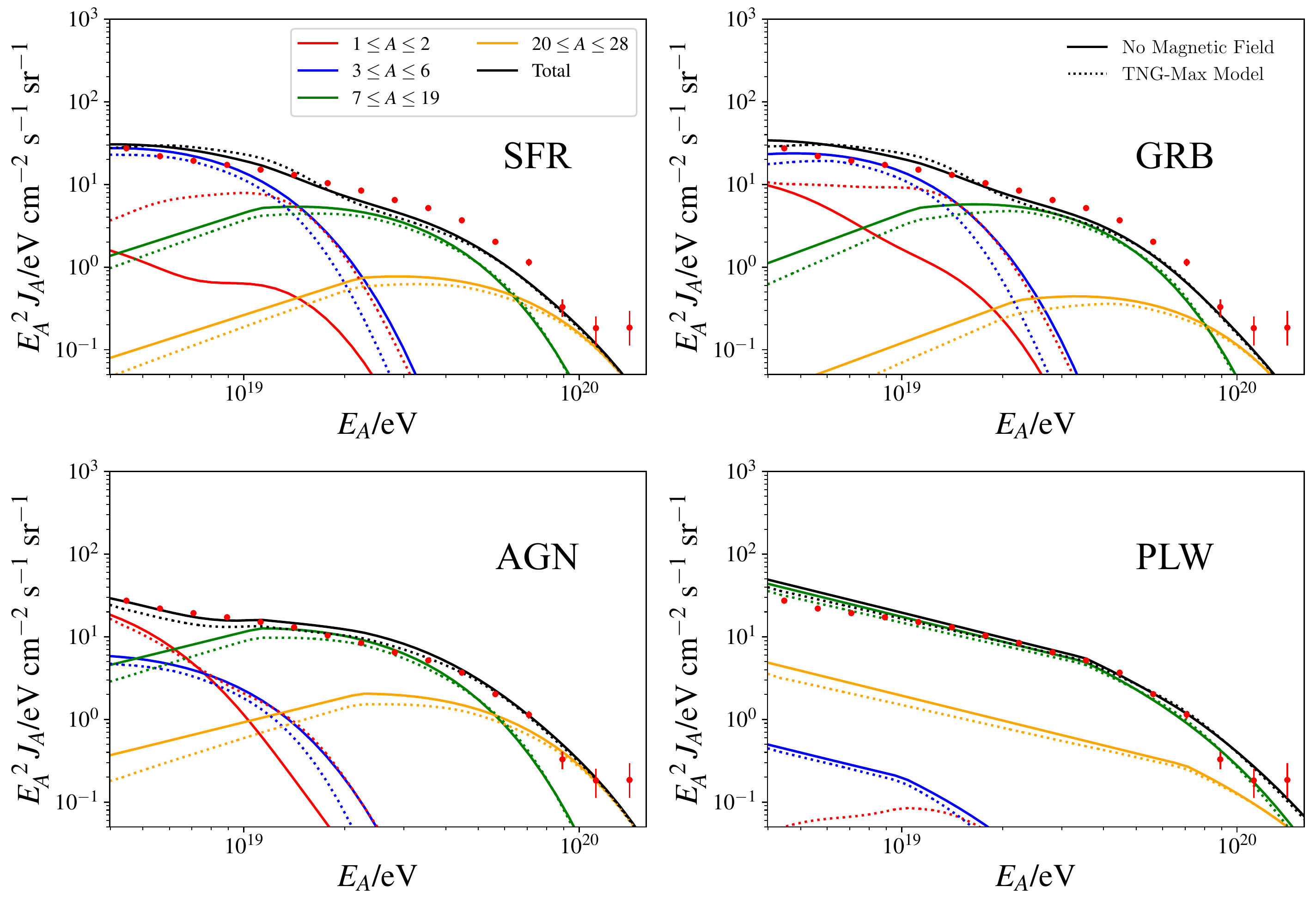}
    \caption{UHE CR flux spectra of the four source population models  
    (SFR, GRB, AGN and PLW) 
    at $z=0$. 
    Solid lines represent scenarios with no IGMFs. Dotted lines represent the scenarios 
     with IGMFs, 
    whose strength evolves according to the TNG-Max prescription (see Section  \ref{sec:magnetic_field_models}). 
    Magnetic horizon effects reduce the overall primary particle flux in the TNG-Max prescription. This is more pronounced in source distributions with heavier compositions, which have shorter magnetic horizon distances. If source distributions are weighted towards higher redshifts and the injection composition is relatively low-mass, secondary production can increase the particle flux at low energies (e.g. as in the SFR and GRB cases).
    The corresponding results 
    for the TNG-Min prescription 
    are not shown. 
    These are very similar to 
     scenarios without IGMFs.  
    Data obtained by PAO \cite{Verzi2019PoS} 
    (red dots with error bars) 
    are shown for comparison. Note that the $1 \leq A \leq 2$ line for the PLW case without magnetic fields falls below the range of the axes. In all models, a further secondary $A=1$ particle component emerges slightly below 10$^{20}$ eV (also below the axes range). These originate from photo-spallation interactions of very energetic heavy CRs with EBL photons. Their contribution is negligible.}
    \label{fig:full_output_results}
\end{figure*}

\begin{figure*}
    \centering
    \includegraphics[width=\textwidth]{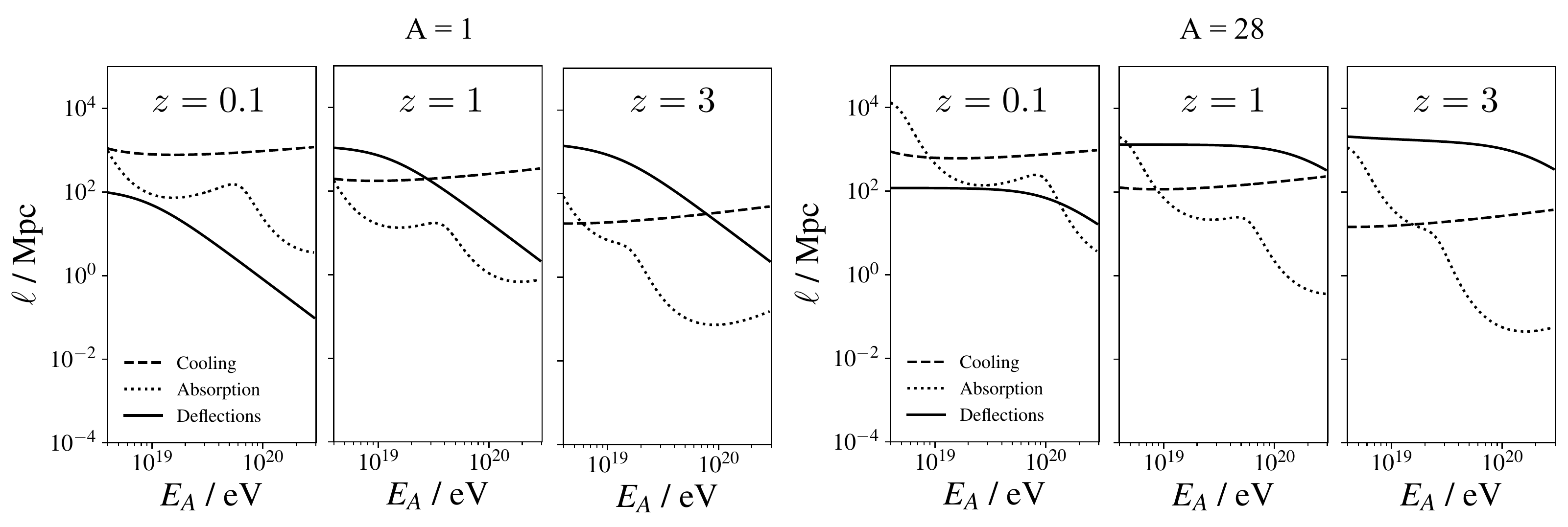}
    \caption{UHE CR cooling (dashed lines), absorption/spallation (dotted lines) distances, and cumulative extra path lengths introduced by deflections (solid lines) in CR propagation scenarios in IGMFs described by the TNG-Max model, from $z=0$ up to $z=0.1$, 1 and 3, as stated. The cumulative extra path lengths only increase significantly up to $z=1$, reflecting the epoch where the evolving IGMF is strongest (see Fig.~\ref{fig:z_evo_magfield}). For $A=1$, absorption is due to photo-pion production. Photo-spallation is the dominant absorption process for heavier nuclei.}
    \label{fig:path_lengths_fig}
\end{figure*}

The UHE CR spectrum at $z_{\rm min} = 0$ for each source class 
  is obtained by 
  integrating the modified transport equation 
(equation~\ref{eq:transport_equation_steadystate_mod}) 
  numerically over a discretized grid in redshift.\footnote{The grid resolution is informed by the shortest interaction path length experienced by the UHE CRs (see Fig.~\ref{fig:path_lengths_fig}). This safeguards against under-predicting attenuation effects and secondary CR production. 
  In this work we have adopted an approach 
   in which the injection and loss terms are approximated 
   analytically, with the average values of 
   interaction rates being used between grid points. 
   Monte Carlo simulations, to be considered in our future studies, 
   will better capture some subtle stochastic aspects 
   in the CR transport, 
   one of such being the evolution 
   (e.g. broadening) of the distribution function 
   of the UHE CR ensemble.}
  The effect of the evolving magnetic field
  is accounted for by calculating the average value of 
   the scaling factor $\Psi_A$ between subsequent grid points. 
  The extra path lengths 
  accounting for CR deflections
  extends the propagation times 
  of the particles.  
In some cases, 
  this 
  may exceed the Hubble time. 
  Such particles will not reach us, forming a magnetic horizon~\cite{Lemoine2005PhRvD, Berezinsky2007ApJ}. 
Their contribution is excluded in the calculations by setting an upper limit in
the integration of $z_{\rm max} = \min \left\{z_{\rm H}, 3 \right\}$, where $z_{\rm H}$ is the redshift at which the total propagation time of a CR undergoing deflections would equal the age of the Universe. This effect changes the total CR arrival flux of each of the species.  
The all-particle spectrum  
  and four broadband composition spectra 
  are shown in Fig.~\ref{fig:full_output_results}. 
For scenarios with no IGMFs, 
  our results are identical 
  to those obtained for streaming CRs. 
If weak IGMFs are present, e.g. TNG-Min (not shown), 
  their effects on CR propagation are insufficient 
  to give results which are noticeably different 
  to scenarios without IGMFs. These results are generally consistent 
 with PAO data \cite{Verzi2019PoS}.  
 
Effects caused by deflections are 
  important in the TNG-Max prescription.  
The results are more difficult 
 to reconcile with the PAO observations \cite{Verzi2019PoS}
 (especially for scenarios invoking SFR or GRB source classes). 
In this situation, strong deflections can become important for some CRs (see Fig.~\ref{fig:path_lengths_fig}). These CRs would propagate via diffusion. The relatively small separation between the sources in our model\footnote{{Separations are estimated to be below $\sim 10$~Mpc for each of the source population models, when adopting individual UHE CR source luminosities of 
$10^{38}-10^{44}$~erg~s$^{-1}$ 
\cite{Alves2019FrASS}.}} is less than the CR diffusion length\footnote{{The diffusion length may be estimated as $\ell_{\rm d} \gtrsim 50$~Mpc, if taken to be at least the distance to last scattering for filaments~\cite{Kotera2008PhRvD}.}} and energy loss distance. Under these conditions, the diffuse average flux spectrum is well approximated by the streaming treatment with deflections adopted here~\cite{Aloisio2004ApJ}. 

The effects of IGMFs 
  on the CR spectrum and composition are shown in Fig.~\ref{fig:full_output_results}. They 
  are manifestations of the competition 
  between cooling and absorption, 
  secondary nuclei production and magnetic horizon effects. 
Qualitatively, 
  we may discern two regimes by CR energy. 
At energies above $\sim 3\times 10^{19}$ eV, 
 deflections introduce an extra CR path length
 of hundreds of Mpc. 
This is longer than 
 photo-spallation lengths, particularly  
  for heavy CR nuclei (see Fig.~\ref{fig:path_lengths_fig}). Cooling lengths at these energies are much longer than attenuation lengths. The high energy spectrum of particles is therefore dominated by the injection of primary CRs, and the effects of cooling are obscured by attenuation.

\begin{figure*}
    \centering
    \includegraphics[width=0.95\textwidth]{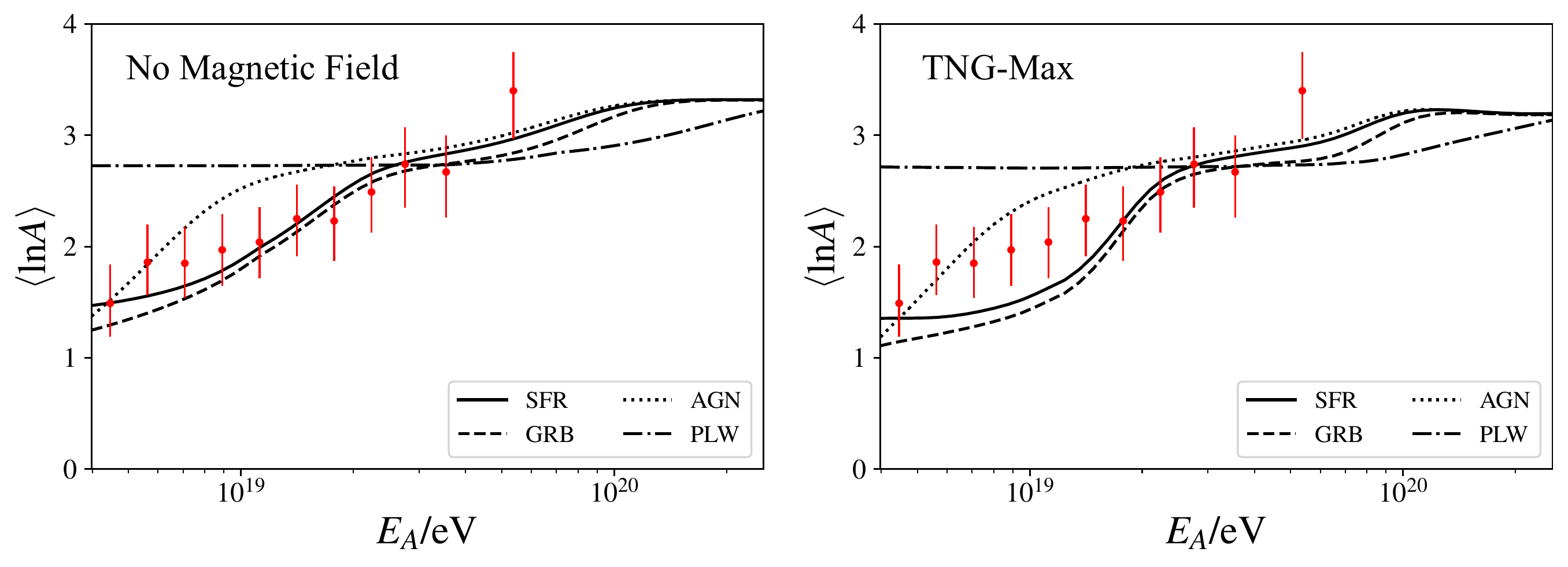}
    \caption{Average UHE CR mass composition $\langle \ln A\rangle$ for the four source classes. The left panel shows scenarios where
    deflections in the IGMFs are not considered. The right panel shows the composition when adopting the TNG-Max prescription. Results obtained with the TNG-Min prescription are indistinguishable from scenarios neglecting IGMFs. The data points shown in red were obtained from PAO~\cite{PAO2013JCAP, Yushkov2019PoS} (using Sibyll 2.3c~\citep{Riehn2017PoS, Fedynitch2019PhRvD}). }
    \label{fig:mean_log_A}
\end{figure*}

While magnetic fields do not  
  have strong effects on the horizon 
  for CRs with extremely high energies,  
the situation is different 
 for CRs with energies 
  below $\sim 3 \times 10^{19}~{\rm eV}$
  where the extended path lengths
  range between tens to hundreds of Mpc.  
This is shorter 
than the 
  photo-pion and photo-spallation length scales 
  at $z<1$ for low mass nucleons, but becomes 
 comparable at higher masses (see Fig.~\ref{fig:path_lengths_fig}). 
   As photo-spallation (for $A>1$) at these energies 
  always dominates over photo-pion absorption, 
  secondary CRs with low mass ($1 \leq A\leq 2$) 
  accumulate.
 At high redshifts, this accumulation 
  is partially countered by photo-pion attenuation and magnetic horizons in the extended path lengths and 
  does not greatly affect the spectrum observed at $z=0$. At lower redshifts, the longer attenuation path lengths mean that the accumulation of secondaries in the presence of IGMFs is more apparent, especially when the injected composition of CRs is of relatively low-mass. As photo-pair lengths are shorter than the extended path lengths and also comparable to attenuation length scales, 
  noticeable cooling effects on the low-mass components of the particle spectrum emerge (in particular, for $A\leq 6$ in the SFR and GRB models). The combination of all these effects in the extended propagation paths experienced by CRs in the presence of IGMFs distort the average mass composition 
  $\langle \ln A \rangle$ at $z=0$, and generally boost the relative abundance of lower-mass nuclei at lower energies (see 
  Fig.~\ref{fig:mean_log_A}).

%
\subsection{Scattering centers and CR source population models}  

When comparing the 
   PAO spectral and composition data in  
  Figs.~\ref{fig:full_output_results} and~\ref{fig:mean_log_A}, scenarios predicting  
  a strong secondary CR component 
  would appear to be less favorable. 
  In the presence of filament magnetic fields 
  of 10s of nG below $z\sim 1$,\footnote{RM (rotation measure) observations reveal strengths of this level are reasonable for cosmic filaments~\citep[see][]{2022MNRASCarretti}.} 
  source distributions weighted 
  towards higher redshifts 
  (e.g. AGN populations as CR sources)
  are preferred. 
 However, source populations (initial conditions) 
  and magnetic field prescriptions 
  (a component in the transport process) 
  are degenerate 
  when matching the results of calculations 
  with observations.  
  
\begin{figure}
    \centering
    \includegraphics[width=\columnwidth]{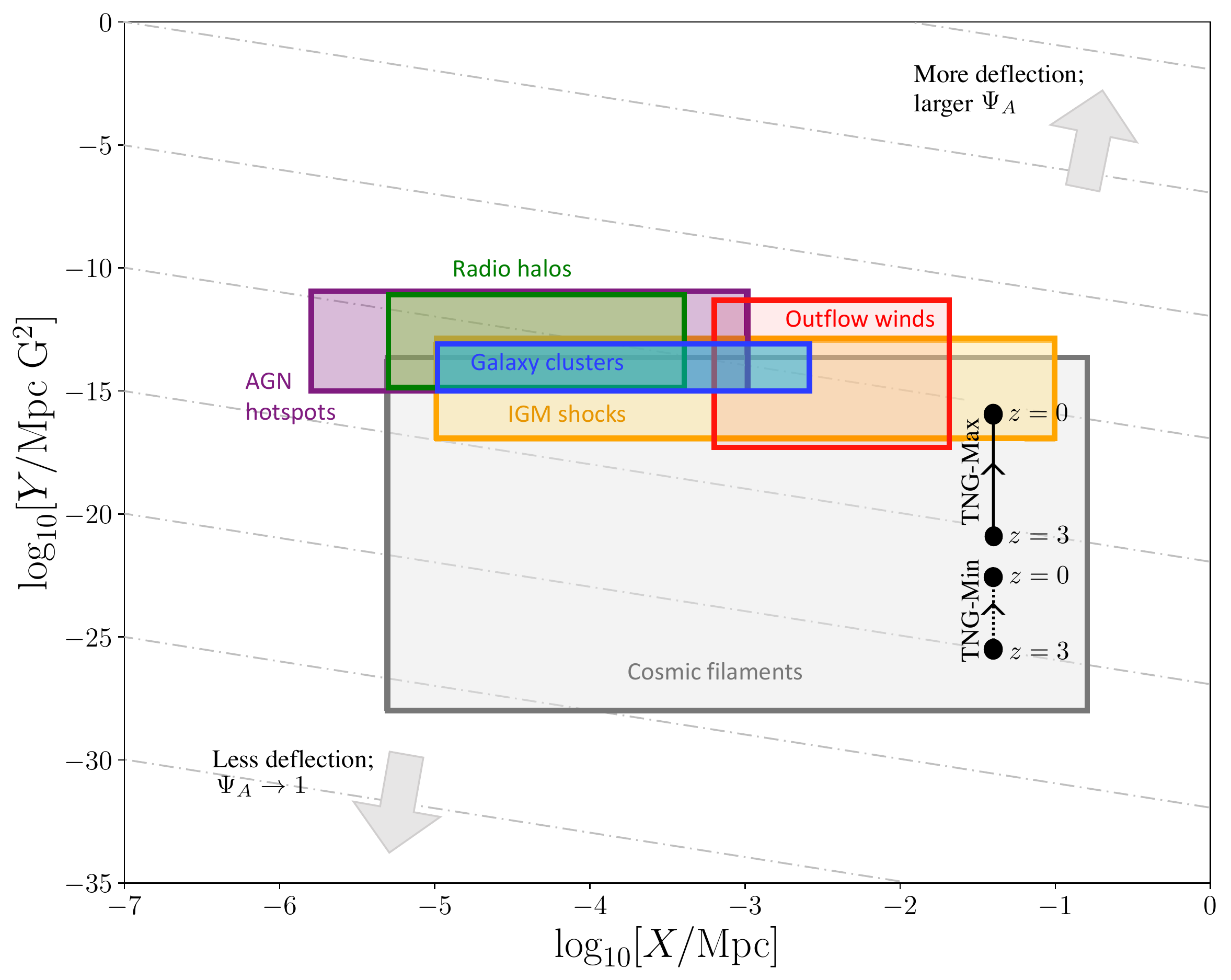}
    \caption{Comparison of 
    the ability of scattering center types to deflect UHE CRs. 
    Here, $X = n_c \;\! {r_c}^2 \;\! \sigma_c$ ($
 \propto n_c \;\! {r_c}^4$) and $Y = {B_c}^2 \;\!\lambda_c$. Regions towards the top right 
    yield stronger deflections. 
    The evolutionary progression of the TNG-Min and TNG-Max prescriptions  
    (see section~\ref{sec:results_flux_comp}) are marked by black arrows. 
    Contours of equal deflection are represented by gray dot-dashed lines.   
Candidate classes of scattering centers 
    are represented by rectangular boxes, 
    where boundaries are set 
    by theoretical or observational constraints 
    (see Appendix~\ref{sec:scattering_center_pops}). }
    \label{fig:mag_scattering_centers}
\end{figure}

In this work, 
  we consider a single type of scattering center -- 
  cosmic filaments. 
  Fig.~\ref{fig:mag_scattering_centers} show 
    also other scattering center candidates  for comparison.  
 Limits bounding each region 
  are set by the approximate range of characteristic sizes, 
  their spatial abundance in the Universe, any knowledge of the magnetic fields inherent to each environment, and observational constraints (obtained from~\cite{Durrer2013AARv, Planck2016AA, Han2017ARAA}; see Appendix~\ref{sec:scattering_center_pops}).
This is presented in terms of two quantities, 
  $X = n_c \;\! r_c^2 \;\! \sigma_c$ ($
 \propto n_c \;\! r_c^4$)
  and $Y = B_c^2 \;\!\lambda_c$, 
  for which the product is proportional to
  the magnetic scaling factor $\Psi_A$ 
  (see equation~\ref{eq:modxx}). 
The top-right 
  of Fig.~\ref{fig:mag_scattering_centers} 
  represents conditions 
  where the deflection potential is stronger.

Figs.~\ref{fig:full_output_results} and~\ref{fig:mean_log_A} 
  show that the effects of IGMFs become substantial 
  at low redshifts 
  for the TNG-Max prescription, 
  and are inconsequential 
  for the TNG-Min prescription.  
The parameter space 
  where the effects of deflections are important 
  lies above a contour 
  through the mid-point of the TNG-Max line  
  in Fig.~\ref{fig:mag_scattering_centers}. 
This passes through 
  the allowed regions of 
  all scattering centers considered. 
Constraints obtained by current observations 
  or theoretical studies 
  are insufficient 
  for distinguishing the merits 
  of the different classes of scattering centers.  
This must be properly addressed    
  when endorsing any UHE CR source class. 
For example, 
   most source populations are acceptable  
   if IGMFs are ignored, 
   but none are acceptable 
   if the TNG-Max prescription is used 
   to derive the IGMF strength and its evolution. 
Nonetheless,  
  Fig.~\ref{fig:mag_scattering_centers} 
  provides useful insights 
  for identifying possible 
  scattering centers, 
  and for modeling their effects on 
  CR transport. 
With up-coming instruments (e.g. the Square Kilometer Array, SKA)\footnote{See: \url{https://www.skatelescope.org}}
  dedicated 
  to study the magnetic properties 
  of the hierarchy of structures 
  in the Universe,  
   more advanced modeling 
  of IGMFs
  will be possible 
 down to galactic scales. 
  With these new insights, 
  transport calculations 
  with a proper astrophysical set up and robust parameter choices  
  will allow us to confidently resolve the origins of UHE CRs.  
  
\section{Summary} 
\label{sec:conclusions}

We investigate the effects of IGMFs 
  on the propagation of UHE CRs 
  and on the CR spectrum and composition  
  observed at $z=0$. 
We solve the particle transport equation 
  accounting for the 
   deflections of CRs by IGMFs, 
   and the cumulative effects of 
   absorption, spallation and interactions 
   with intergalactic radiation fields. 
Piece-wise deflections of CR particles by IGMFs 
  are modeled  
  as stochastic scattering by centers 
  associated with cosmic filaments. 
The properties of the filaments 
  and the evolution of their magnetic field strengths  
  are derived from cosmological simulations.
Our calculations  
  have shown that IGMFs 
  can have a marked effect 
  on the observed properties of UHE CRs 
  detected on Earth, 
  when the IGMF strength reaches 
  10s of nG 
  for cosmic filaments by $z\lsim 1$.  
We find the source population models  
  and IGMFs are degenerate. 
This degeneracy   
  must be properly resolved 
  before endorsing or disfavoring  
  different UHE CR source classes 
  (including those not considered in this work) 
  to determine the origin of components 
  in broad composition spectra  
  of UHE CRs observed on Earth. 
Refinement of the results obtained in this work 
  can be achieved 
  by improving the modeling 
  of the IGMFs in cosmic filaments 
  and their substructures.   
 Observations 
  by near-future facilities, in particular, the SKA,
  will advance our knowledge 
  of the hierarchical properties 
  of magnetic fields from scales of 
  galaxies and clusters to filaments and voids, 
  thus providing more robust inputs  
  for modeling the scattering of UHE CRs by IGMFs  
  in transport calculations.

\vspace{0.2cm}
\textbf{Acknowledgments}: 
We thank the anonymous referees for their constructive comments. 
ERO is an overseas researcher under the Postdoctoral Fellowship of the Japan Society for the Promotion
of Science (JSPS), supported by JSPS KAKENHI Grant Number JP22F22327, 
and also acknowledges support from the Center for Informatics and Computation in Astronomy (CICA) at National Tsing Hua University (NTHU) through a grant from the Ministry of Education (MoE) of Taiwan (ROC), where part of this study was conducted. 
QH is supported by a UCL Overseas Research Scholarship 
  and a UK Science and Technology Facilities Council (STFC) Research Studentship. 
QH and KW acknowledge support from 
  the UCL Cosmo-particle Initiative. 
This study is supported in part 
  by a UK STFC Consolidated Grant awarded to UCL-MSSL. 
This work made use of high-performance computing facilities operated by CICA at NTHU. 
This equipment was funded by the MoE and the National Science and Technology Council of Taiwan (ROC). 
This work made use of IllustrisTNG simulations, which were undertaken with compute time awarded by the Gauss Centre for Supercomputing (GCS) under GCS Large-Scale Projects GCS-ILLU and GCS-DWAR on the GCS share of the supercomputer Hazel Hen at the High Performance Computing Center Stuttgart (HLRS), as well as on the machines of the Max Planck Computing and Data Facility (MPCDF) in Garching, Germany. 
This work made use of the NASA ADS database. 

\appendix

\section{Scattering centers}
\label{sec:scattering_center_pops}

\begin{table*}
\begin{adjustbox}{width=\textwidth,center}
\begin{tabular}{|l|lr|lr|lr|lr|}
 \hline
 \multirow{2}{*}{Type} & \multicolumn{8}{c|}{Parameter value range} \\
 & \multicolumn{2}{c|}{$n_c$/Mpc$^{-3}$} & \multicolumn{2}{c|}{$r_c$/Mpc} & \multicolumn{2}{c|}{$\lambda_c$/Mpc} & \multicolumn{2}{c|}{$B_c$/$\mu$G} \\
  \hline
 Cosmic filaments & $(0.3-1) \times 10^{-2}$ & \textit{See note}$^{(a)}$ & $0.2 - 2$ & \textit{Ref. \cite{Gheller2019MNRAS}} & \multicolumn{2}{c|}{\textit{See note}$^{(b)}$} & $10^{-5.5} - 10^{-1}$ & \textit{Ref. \cite{Marinacci2018MNRAS}} \\
 Galaxy clusters & $10^{-5}$ & \textit{Ref.~\cite{Kotera2008PhRvD, Oswalt2013pss6}} & $1 - 4$ & \textit{Ref.~\cite{Hansen2005ApJ}} & 0.1 & \textit{See note}$^{(c)}$ & $10^{-1} - 10^{0}$ & \textit{Ref.~\cite{Bruggen2005ApJ, Vacca2018Galax}} \\
 Galactic outflow winds & $(1-5) \times 10^{-2}$ & \textit{Ref. \cite{Kotera2008PhRvD, Bianconi2020MNRAS}} & $0.5 - 0.8$ & \textit{Ref. \cite{Kotera2008PhRvD}} & 0.05 & \textit{Ref.~\cite{Kotera2008PhRvD}} & $10^{-2} - 10^{1}$ & \textit{Ref. \cite{Heesen2016MNRAS, Pakmor2020MNRAS}} \\
 Radio halos & $5\times 10^{-6}$ & \textit{Ref. \cite{Cuciti2015AA}}$^{(d)}$ & $1 - 3$ & \textit{Ref. \cite{Xie2020AA, Hoang2021AA}} & 0.1 & \textit{See note}$^{(c)}$ & $10^{-1} - 10^{1}$ & \textit{Ref. \cite{Kotera2008PhRvD, Stanev2010hecr}} \\
 IGM accretion shocks around clusters & \textit{$10^{-5}$} & \textit{See note}$^{(e)}$ & $1 - 10$ & \textit{Ref. \cite{Kotera2011ARAA}} & 0.1 & \textit{See note}$^{(c)}$ & $10^{-2} - 10^{0}$ & \textit{Ref. \cite{Weeren2010Sci}} \\
 AGN terminating bow shocks (hot spots) & $10^{-3}$ & \textit{Ref.~\cite{Silverman2005ApJ}}$^{(f)}$ & $0.2 - 1$ & \textit{Ref. \cite{Stanev2010hecr, Orienti2012MNRAS}} & 0.1 & \textit{See note}$^{(c)}$ & $10^{-1} - 10^{1}$ & \textit{Ref. \cite{Werner2012ApJ, Stanev2010hecr}} \\
 \hline
  \end{tabular}
  \end{adjustbox}
 \label{tab:scat_cent_params}
\caption{Scattering centers parameters, 
  as in Fig.~\ref{fig:mag_scattering_centers}. \textbf{Notes}: {\bf (a)} Values estimated from the sizes of voids separating filaments, combined with the indicated $r_c$ range (to estimate a filament cross section). {\bf (b)} This value is not specified. Constraints of $\lambda_c$ are derived from those in~\cite{Durrer2013AARv,Planck2016AA,Han2017ARAA} based on adopted $B_c$ values, where $\lambda_c$ has an upper limit of $r_c$. {\bf (c)} Default value is used for IGMFs (see, e.g.~\cite{Ryu2008Sci}). {\bf (d)} Based on~\cite{Cuciti2015AA}, we assume that 50\% of clusters host radio halos. 
  This fraction depends on cluster mass and 
  merging history. {\bf (e)} We assume this to be the same as for galaxy clusters, where accretion shocks are often present. {\bf (f)} Estimated from X-ray selected AGN, many of which tend to have hot spots.}
\end{table*}

The ranges of parameter values adopted 
for each type of scattering center shown in Fig.~\ref{fig:mag_scattering_centers} are summarized in Table II.

\bibliographystyle{h-physrev}
\bibliography{reference}

\end{document}